\documentclass[%
superscriptaddress,
 preprint,
 amsmath,amssymb,
 floatfix,
]{revtex4-2}

\usepackage{booktabs}
\usepackage{array}
\newcolumntype{L}[1]{>{\raggedleft\arraybackslash}p{#1}}
\AtBeginDocument{
\heavyrulewidth=.08em
\lightrulewidth=.05em
\cmidrulewidth=.03em
\belowrulesep=.65ex
\belowbottomsep=0pt
\aboverulesep=.4ex
\abovetopsep=0pt
\cmidrulesep=\doublerulesep
\cmidrulekern=.5em
\defaultaddspace=.5em
} 

\usepackage{setspace} % Load the setspace package
\usepackage{soul}
\usepackage{hyperref}
\usepackage{color}
\usepackage{graphicx}% Include figure files
\usepackage{dcolumn}% Align table columns on decimal point
\usepackage{bm}% bold math
\usepackage{cancel}
\usepackage[utf8]{inputenc}

\usepackage{upgreek}
\usepackage[normalem]{ulem}
\newcommand{\ie}{i.e.}

\newcommand{\citar}[1]{{\color{blue}[CITE]}}
\begin{document}

\title{Unidirectional motion of topological defects mediating continuous rotation processes}

\author{Marisel Di Pietro Martínez}
\email{Marisel.DiPietro@cpfs.mpg.de}
\affiliation{Max Planck Institute for Chemical Physics of Solids, Noethnitzer Str. 40, 01187 Dresden, Germany}
\affiliation{International Institute for Sustainability with Knotted Chiral Meta Matter (WPI-SKCM$^2$), Hiroshima University, Hiroshima 739-8526, Japan}
\author{Luke Alexander Turnbull}
\affiliation{Max Planck Institute for Chemical Physics of Solids, Noethnitzer Str. 40, 01187 Dresden, Germany}
\affiliation{International Institute for Sustainability with Knotted Chiral Meta Matter (WPI-SKCM$^2$), Hiroshima University, Hiroshima 739-8526, Japan}
\author{Jeffrey Neethirajan}
\affiliation{Max Planck Institute for Chemical Physics of Solids, Noethnitzer Str. 40, 01187 Dresden, Germany}
\author{Max Birch}
\affiliation{RIKEN Center for Emergent Matter Science (CEMS) Wako 351-0198, Japan}
\author{Simone Finizio}
\author{J\"org Raabe}
\affiliation{Swiss Light Source, Paul Scherrer Institut, Forschungsstrasse 111 5232 PSI Villigen, Switzerland}
\author{Edouard Lesne}
\affiliation{Max Planck Institute for Chemical Physics of Solids, Noethnitzer Str. 40, 01187 Dresden, Germany}
\author{Anastasios Markou}
\affiliation{Max Planck Institute for Chemical Physics of Solids, Noethnitzer Str. 40, 01187 Dresden, Germany}
\affiliation{Physics Department, University of Ioannina, 45110 Ioannina, Greece}
\author{María Vélez}
\author{Aurelio Hierro-Rodríguez}
\affiliation{Departamento de Física, Universidad de Oviedo, 33007, Oviedo, Spain}
\affiliation{CINN (CSIC-Universidad de Oviedo), 33940, El Entrego, Spain}
\author{Marco Salvalaglio}
\affiliation{Institute of Scientific Computing, TU Dresden, 01062 Dresden, Germany}
\affiliation{Dresden Center for Computational Materials Science (DCMS), TU Dresden, 01062 Dresden, Germany}
\author{Claire Donnelly}
\email{claire.donnelly@cpfs.mpg.de}
\affiliation{Max Planck Institute for Chemical Physics of Solids, Noethnitzer Str. 40, 01187 Dresden, Germany}
\affiliation{International Institute for Sustainability with Knotted Chiral Meta Matter (WPI-SKCM$^2$), Hiroshima University, Hiroshima 739-8526, Japan}

\date{\today}% It is always \today, today,
             % but any date may be explicitly specified

\begin{abstract}
% <= 200 words
\textbf{Topological defects play a critical role across many fields, mediating phase transitions and macroscopic behaviors as they move through space. Their role as robust information carriers has also generated much attention. However, controlling their motion remains challenging, especially towards achieving motion along well-defined paths which typically require predefined structural patterning.
Here we demonstrate the tunable, unidirectional motion of topological defects, specifically magnetic dislocations in a weak magnetic stripe pattern, induced by external magnetic field in a laterally unconfined thin film. This motion is shown to mediate the overall continuous rotation of the stripe pattern. We determine the connection between the unidirectional motion of dislocations and the underlying three-dimensional (3D) magnetic structure by performing 3D magnetic vectorial imaging with \textit{in situ} magnetic fields. A minimal model for dislocations in stripe patterns that encodes the symmetry breaking induced by the external magnetic field reproduces the motion of dislocations that facilitate the 2D rotation of the stripes, highlighting the universality of the phenomenon. This work establishes a framework for studying the field-driven behavior of topological textures and designing materials that enable well defined, controlled motion of defects in unconfined systems, paving the way to manipulate information carriers in higher-dimensional systems.}
\end{abstract}

\maketitle
% less than 3000 words

%##################################################################################
% \section{Introduction}

Topological defects~\cite{mermin1979topological} are ubiquitous across various physical systems. From vortex-antivortex pairs in superconductors~\cite{langer1967intrinsic} to dislocations in crystals~\cite{anderson2017,harrison2000mechanisms,harrison2002dynamics}, topological defects emerge even in active matter~\cite{shankar2022topological}, where in contexts as biology, they are found to mediate fundamental processes such as tissue regeneration and cell death~\cite{ardavseva2022topological}.
In magnetism, topological defects not only mediate processes that influence the macroscopic behavior of materials but, at the nanoscale, they can be used to carry information~\cite{parkin2008magnetic} - opening the door to fascinating fundamental physical phenomena and technological applications. Indeed, topological defects can be considered to have a double role: as mediators of processes as well as information carriers.

As mediators, topological defects can be studied as quasi-particles that can be created and destroyed, and that move through space~\cite{dussaux2016local}.
For instance, controlling the creation of structural dislocations can enhance the performance of non-rare-earth permanent magnets~\cite{jia20201} used in technology and energy generation.
Another example is magnetic dislocations in helimagnets, which facilitate skyrmion and bubble nucleation under a magnetic field, mediating the transition between stripe and skyrmionic phases~\cite{yu2012magnetic}.
To control these mediated processes, it is crucial to understand how topological defects respond to external excitations like magnetic fields or electric currents.

On the other hand, as information carriers, topological defects have attracted recent attention since they can be propagated and controlled, and represent an energy-efficient alternative to conventional computing devices.
For instance, in magnetic racetracks, information is carried by domain walls as they move along a patterned nanostrip~\cite{parkin2008magnetic}.
This one dimensional (1D) motion is achieved by the geometric confinement of the track, enabling precise control over the transfer of information from one point to another.
Going beyond confined systems promises higher complexity, density and reconfigurability needed for non-traditional computing and logic devices~\cite{raab2022brownian}. However, achieving such controlled motion of defects in unconfined systems can be more challenging.
In the case of skyrmions, for instance, it requires the cancellation of the skyrmion Hall effect to move in straight lines within two-dimensional (2D) planes~\cite{litzius2017skyrmion,jiang2017direct}.
Indeed, when realizing topological defects as information carriers, the main challenge resides in achieving precise control over their motion, raising the question of whether the 1D motion of defects in an unconfined system is possible.
Recently, unidirectional motion of defects has been achieved without the need for nanopatterning of materials - but rather by promoting the ``gliding'' of defects along the direction of straight stripe domain patterns~\cite{hierro2017deterministic,fernandez2024memory,he2024experimental}.
Although in those cases the motion is restricted to the stripe direction, this prospect of achieving controllable, unidirectional motion in unconfined systems offers new possibilities for future information transport devices.

Here we present a laterally unconfined system where topological defects exhibit a well-defined lateral 1D motion, the direction of which can be chosen
within the two dimensional plane, mediating a continuous transition of the system order parameter.
The system consists of a magnetic thin film hosting weak stripe domains that rotate continuously when subjected to an external in-plane magnetic field.
Intriguingly, the 1D motion is not confined to the direction of the stripe background - but occurs in the 2D plane of the system and it is defined by the combination of the stripe orientation and the magnetic field.
%, mediating the continuous rotation of the overall stripe pattern under the application of a magnetic field.
We demonstrate that the continuous behavior of the rotation transition is enabled by topological defects, specifically magnetic dislocations, which propagate in a well-defined 1D direction despite the unconfined nature of the film.
The direction of motion of these defects can be controlled using a magnetic field, opening the possibility to propagate defects along arbitrary paths in 2D planes.
To determine the mechanism behind the motion of the dislocations, we perform high-resolution magnetic imaging with synchrotron X-rays, revealing the multidimensional nature of the system: the one dimensional motion of the magnetic defects mediates a two dimensional rotation process, that relies on the underlying three dimensional (3D) magnetic structure.
Furthermore, we reproduce the unidirectional motion of defects driving the rotation transition using a minimal model that generally describes an ordered stripe pattern containing defects. This minimal model is relevant beyond magnetism, thus highlighting the universality of this phenomenon.
Altogether, the understanding gained with this multidimensional approach opens the door to the design of controlled motion of topological defects within unconfined systems.

\begin{figure*}[!t]
\begin{center}
\includegraphics[width=\linewidth]{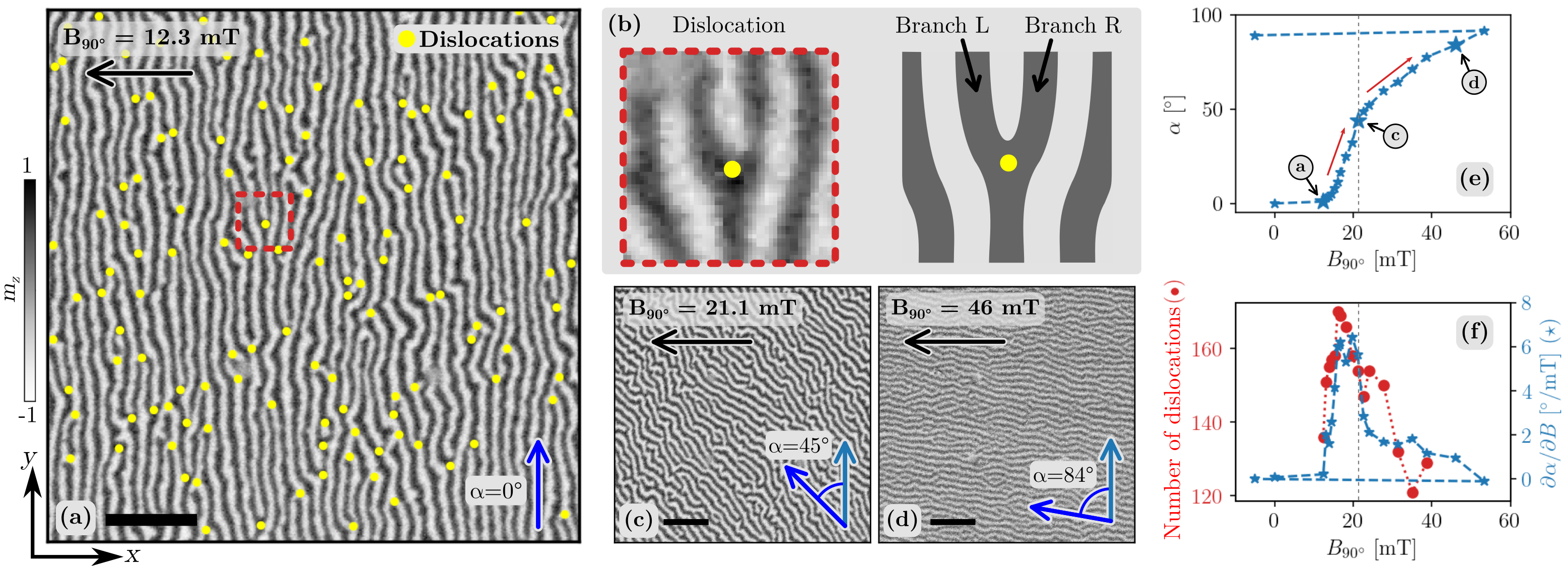}
\end{center}
\caption{Magnetic stripes and dislocations:
(a,c,d) The projection of the out-of-plane magnetization $m_z$ in a $400$\,nm-thick Py film is imaged across a $15\,\upmu$m$ \times 15\,\upmu$m field of view by scanning transmission X-ray microscopy.
The average orientation of the stripes $\alpha$ (blue arrow) rotates as an external magnetic field $B$ (black arrow) is applied perpendicular to the initial orientation of the stripes.
Scale bar represents $2.5\,\upmu$m.
(b) Magnetic dislocations appear when the stripes bifurcate into a left (L) and a right (R) branch.
More than a hundred magnetic dislocations (yellow dots) are present in the system (a).
(e) The average orientation of the stripes $\alpha$ increases continuously with the field $B$.
(f) Comparison between the derivative of the angle of the stripes $\alpha$ with respect to the magnetic field $B$ (blue stars) - \ie{} the rate of rotation - and the number of dislocations (red dots). The peak $\partial\alpha/\partial B$ matches the maximum in the number of dislocations, suggesting a connection between the rotation and the defects.}
% "Upon increasing the applied field, My (red curve) remains practically constant for a certain field interval (up to a certain threshold), whereas Mx (blue curve) linearly increases, with a change of slope when My starts decreasing. The same (scalar) behavior has been experimentally observed and numerically
% simulated in [22] for Fe-Ga films and in [33] for Fe-N thin films" Coisson 2019
\label{fig:fig1}
\end{figure*}

% In the following, we solve
% \begin{itemize}
%     \item What is the role of dislocations in the rotation transformation?
%     \item Why do they move in one direction?
%     \item What is the mechanism for the propagation of the dislocations?
% \end{itemize}

\section{Role of dislocations in the rotation process}

To investigate topological defects as mediators of processes, we select a model system exhibiting a spontaneous symmetry breaking resulting in a stripe pattern.
While this stripe pattern is found in various contexts, including surface wrinkles~\cite{ohzono2006defect} and diblock copolymers~\cite{hur2018defect}, here we focus on stripes occurring in magnetic materials (see Fig.~\ref{fig:fig1} (a)).
In all of these systems, defects known as dislocations occur when stripes bifurcate into two branches (Fig.~\ref{fig:fig1} (b)).
Here we focus on weak magnetic stripes, where elongated domains of canted out-of-plane magnetization form, with a net in-plane magnetic moment.
When imaging the magnetic stripe pattern, we find more than one hundred dislocations in the $15\,\upmu$m$ \times 15\,\upmu$m field of view, highlighted with yellow dots in Fig.~\ref{fig:fig1} (a), making this system ideal to study how these topological defects move and influence the magnetic behavior.

This system not only exhibits a high density of defects but also allows the manipulation of the stripes' orientation through an external in-plane field, a property observed in several magnetic materials~\cite{saito1964new,tee2013magnetization,barturen2012crossover,barturen2013,lehrer1963rotatable,tacchi2014,fin2015,coisson2019,sallica2010magnetic,hierro2020} known as rotatable anisotropy~\cite{prosen1961rotatable,lommel1962rotatable}.
To visualize the weak stripe pattern, we probe the out-of-plane magnetization component in a 400 nm-thick permalloy (Py) film using scanning transmission X-ray microscopy (STXM) with X-ray magnetic circular dichroism.
The sample is prepared by initially applying a saturating vertical ($y$) magnetic field that aligns the stripes vertically.
Subsequently, we apply a horizontal ($x$) magnetic field $B$ at $90^\circ$ to the stripe direction, causing the stripes to rotate, as demonstrated in Figs.~\ref{fig:fig1} (c) and (d).
As the magnetic field $B$ increases, we observe a continuous transition in the average stripe orientation $\alpha$ (see Fig.~\ref{fig:fig1} (e)).
Since the orientation of the stripes remains stable at remanence, the continuous character in the transition suggests that non-volatile analogue information could be potentially encoded in $\alpha$.

A correlation between the rotation of the stripes and the behavior of dislocations is observed when comparing the derivative of the angle of the stripes $\alpha$ with respect to the magnetic field $B$ (blue) to the number of dislocations (red). This effectively compares the rate of rotation to the dislocation population, and is shown in Fig.~\ref{fig:fig1} (f).
The number of dislocations indeed does not remain constant with field but rather increases to a maximum before decreasing again, indicating that dislocations are created and annihilated during the rotation process.
Notably, both the number of dislocations and the rate of rotation peak at the same field, meaning that the stripes rotate at a higher rate when the number of dislocations is higher.
When we consider the rate of rotation, we can identify two regimes: below $B<21.1$\,mT, as the dislocations are created, the rotation rate increases, which is translated into a steeper slope for $\alpha$ vs $B$, while above $B>21.1$\,mT the predominant annihilation of defects reduces the rotation rate, resulting in a lower slope (Fig.~\ref{fig:fig1} (e)).

\begin{figure*}[!t]
\begin{center}
\includegraphics[width=0.98\linewidth]{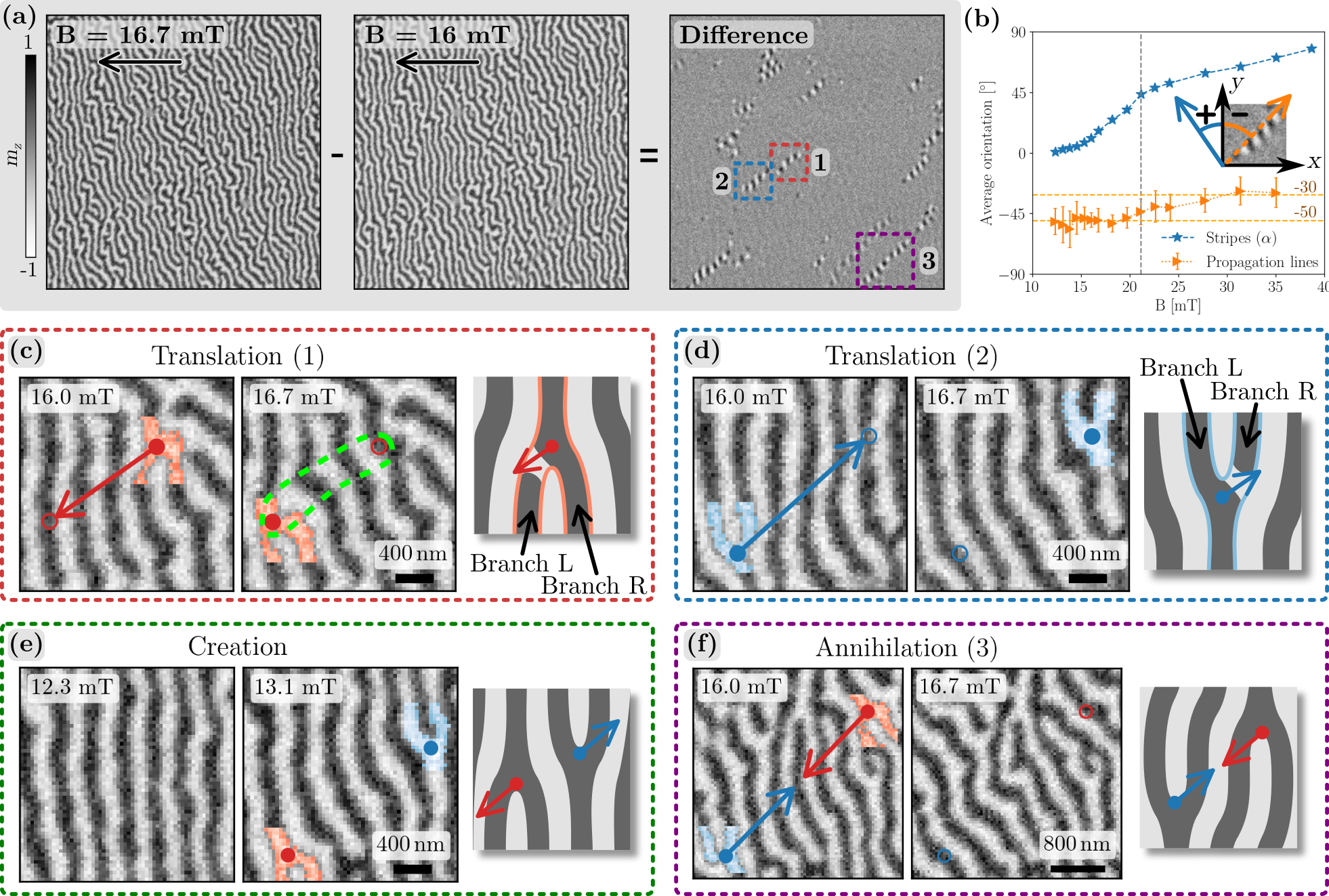}
\end{center}
\caption{Propagation of the defects with magnetic field: (a) The path taken by the defects can be detected by subtracting two images of consecutive field steps.
(b) The average orientation of these lines is approximately constant for fields below $21.1$\,mT (grey vertical line) regardless of the average orientation of the stripes, and increases from $-50(4)$\textdegree{} to $-30(4)$\textdegree{} for higher fields.
Four different types of events are identified: (c) A dislocation with its two branches extending down moves diagonally with the horizontal component of its motion parallel to the magnetic field, acting as a positive particle.
(d) A dislocation with its two branches extending up moves diagonally with the horizontal component of its motion antiparallel to the magnetic field, acting as a negative particle. (e) A pair of positive and negative dislocations are created. (f) A pair of positive and negative particles annihilate each other.}
\label{fig:fig2}
\end{figure*}

To delve deeper into the role of defects in the rotation process, we highlight changes in the stripe configuration by subtracting images measured at two consecutive fields as shown in Fig.~\ref{fig:fig2} (a).
The resulting difference image exhibits zero contrast except along alternating black and white lines that are predominantly oriented in one direction.
The orientation of these lines, plotted against the applied field in Fig.~\ref{fig:fig2} (b), averages a constant $-50$\textdegree{}$\pm 4$\textdegree{} relative to the initial stripe orientation at lower fields and then increases to $-30$\textdegree{}$\pm4$\textdegree{} at higher fields. Performing the experiment for different in-plane orientations yields consistent angular behavior, indicating that the system is isotropic.
Remarkably, these lines are not aligned with the average orientation of the stripes, and therefore indicate the presence of 1D phenomena occurring within the unconfined 2D plane of the system.
The lines are narrower and better defined at lower fields, and become less well defined at higher fields, indicating the presence of multiple regimes of the phenomenon (see Extended Data for all difference images).

To understand the origin of these lines, we compare the two magnetic configurations in the vicinity of the lines.
For line $1$ (green box in Fig.~\ref{fig:fig2} (a)), a dislocation (shaded in red in Fig.~\ref{fig:fig2} (c)) moves diagonally from one end of the line to the other, with the horizontal component of its motion parallel to the magnetic field, acting as a ``positive'' particle.
In contrast, for line $2$ (red box), a dislocation (shaded in blue in Fig.~\ref{fig:fig2} (d)) moves in the opposite direction, hence behaving as a ``negative'' particle.
Dislocations behave as positive or negative depending on the direction of bifurcation, regardless of their net out-of-plane magnetization (black or white contrast). They exhibit indeed a well-defined 1D motion, as suggested by the difference images.
Remarkably, the motion orientation is not solely defined by the orientation of the stripes, but instead by the combination of the field direction with respect to the stripes, as seen in Fig.~\ref{fig:fig2} (b) where the direction of motion is constant despite the changing orientation of the stripes.
This would allow one to propagate the defects along different directions with respect to the stripe orientation, opening the possibility to propagate defects along arbitrary paths in 2D planes controlled by external stimuli.

Moreover, the alternating contrast in these difference images indicates that dislocations oscillate between black and white states, requiring one branch to break for a dislocation to switch states and therefore move.
This diagonal motion combines a parallel (climbing) and perpendicular (gliding/slipping) movement with respect to the stripes. 
Such motion combining climbing and gliding has previously been observed occurring stochastically in magnetic materials~\cite{dussaux2016local,pamyatnykh2017motion}, and represents a universal mechanism for dislocations in general to move in different directions~\cite{anderson2017}. However, this is the first time that such combined motion has been observed following a deterministic well-defined 1D trajectory while mediating a continuous process such as the rotation of the stripes.

As well as observing the motion of dislocations in the system, we also find that these dislocations, akin to charged particles, are also created and annihilated in pairs.
For instance, in Fig.~\ref{fig:fig2} (e), a pair of positive and negative defects emerges from a previously uniform stripe phase when the field increases,
while in Fig.~\ref{fig:fig2} (f), a pair annihilates in the next field step.
These observations demonstrate that the dislocations undergo processes of creation, annihilation and propagation during the rotation of the stripes, thus exhibiting particle/ antiparticle behaviour.

To determine how the movement of defects relates to the overall rotation of the stripes, we examine the local orientation of the stripes in the vicinity of the propagation lines.
In Fig.~\ref{fig:fig2} (c), the area within the dashed-green frame is notably the only region where the stripes have rotated with respect to the previous orientation.
This behavior is consistent across other propagation lines, indicating that dislocation movement drives local stripe rotation, leading to a continuous global reorientation. This mechanism resembles that of dislocation motion known in crystals where the motion of dislocations promotes the local reorientation of the crystal lattice~\cite{wang2014grain}.

Dislocations are crucial mediators in the rotation process, moving in a well-defined direction even without geometric constraints, particularly in the lower field regime. Their unidirectional motion in a two-dimensional system
% , resembling fracton behavior~\cite{pretko2018fracton},
suggests potential as information carriers. However, understanding the underlying mechanism behind this unidirectional movement is essential to harness this phenomenon.

%%%%%%%%%%%%%%%%%%%%%%%%
\section{Effect of the three-dimensional magnetic configuration}

\begin{figure*}[!bt]
\begin{center}
\includegraphics[width=\linewidth]{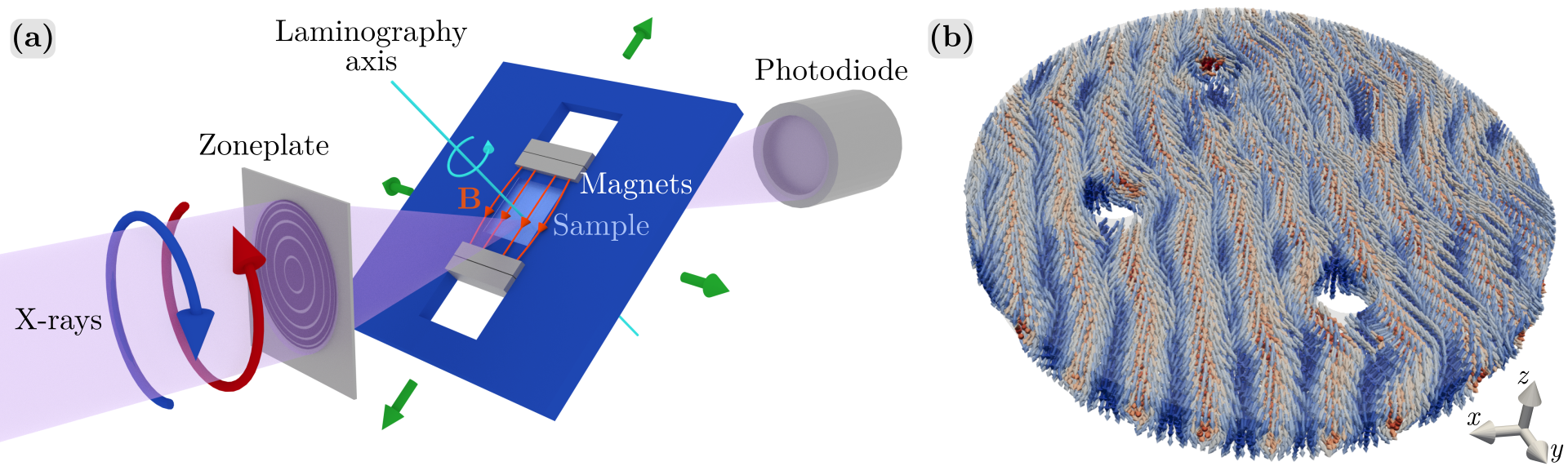}
\end{center}
\caption{3D vectorial magnetic imaging with \textit{in situ} magnetic fields: (a) Circularly polarized X-rays are focused on the sample surface which is tilted at $45$\textdegree{} with respect to the beam, and the transmitted light is measured by a photodiode. The sample is moved in the direction of the green arrows to perform scanning transmission X-ray microscopy (STXM). By rotating the sample around the laminography axis, which corresponds to the surface normal of the sample, a set of 2D STXM images is obtained that can be used to reconstruct the 3D vectorial magnetization. To measure laminography while applying \textit{in situ} magnetic fields, a special sample holder with stacking permanent magnets is used. This design ensures that the magnetic field follows the sample as it rotates, maintaining the same magnetic configuration during the laminography measurement. (b) As a result of the laminography measurement, a 3D vectorial magnetic image is obtained showing the magnetic configuration in a $5\,\upmu$m-diameter cylindrical field of view of the sample. Three holes are milled into the sample to facilitate normalization of the illumination and alignment of the image set.}
\label{fig:fig3}
\end{figure*}

To determine the underlying mechanism for the movement of the dislocations, we consider the multidimensional nature of the system.
While in the previous section only the out-of-plane magnetization component was probed, the stripe domains exhibit a more complex 3D magnetic structure~\cite{hierro2020}.
To determine this 3D configuration under the application of magnetic fields, we perform X-ray magnetic laminography\cite{donnelly2020time,witte2020}, as illustrated in Fig.~\ref{fig:fig3} (a), to obtain full 3D vectorial reconstructions of the magnetisation configuration of the sample (Fig.~\ref{fig:fig3} (b)). While 3D X-ray magnetic imaging is now well established~\cite{donnelly2017three,hierro20183d,di2023three,rana2023}, the incorporation of stimuli is challenging due to the fact that the environment must rotate with the sample. 
Previous studies demonstrate dynamic imaging of the magnetic response to weak magnetic fields~\cite{donnelly2020time}, while insight into the 3D morphology of skyrmion strings under fields has been gained with scalar 3D imaging of a single magnetic component~\cite{seki2022direct}. However, tracking the 3D vectorial configuration under the application of larger magnetic fields has not been possible.
Here, we develop a dedicated sample holder (Fig.~\ref{fig:fig3} (a)) with which we can combine X-ray magnetic laminography with \textit{in situ} magnetic fields to follow the evolution of the 3D configuration of the stripes and defects.

\begin{figure*}[!t]
\begin{center}
\includegraphics[width=0.93\linewidth]{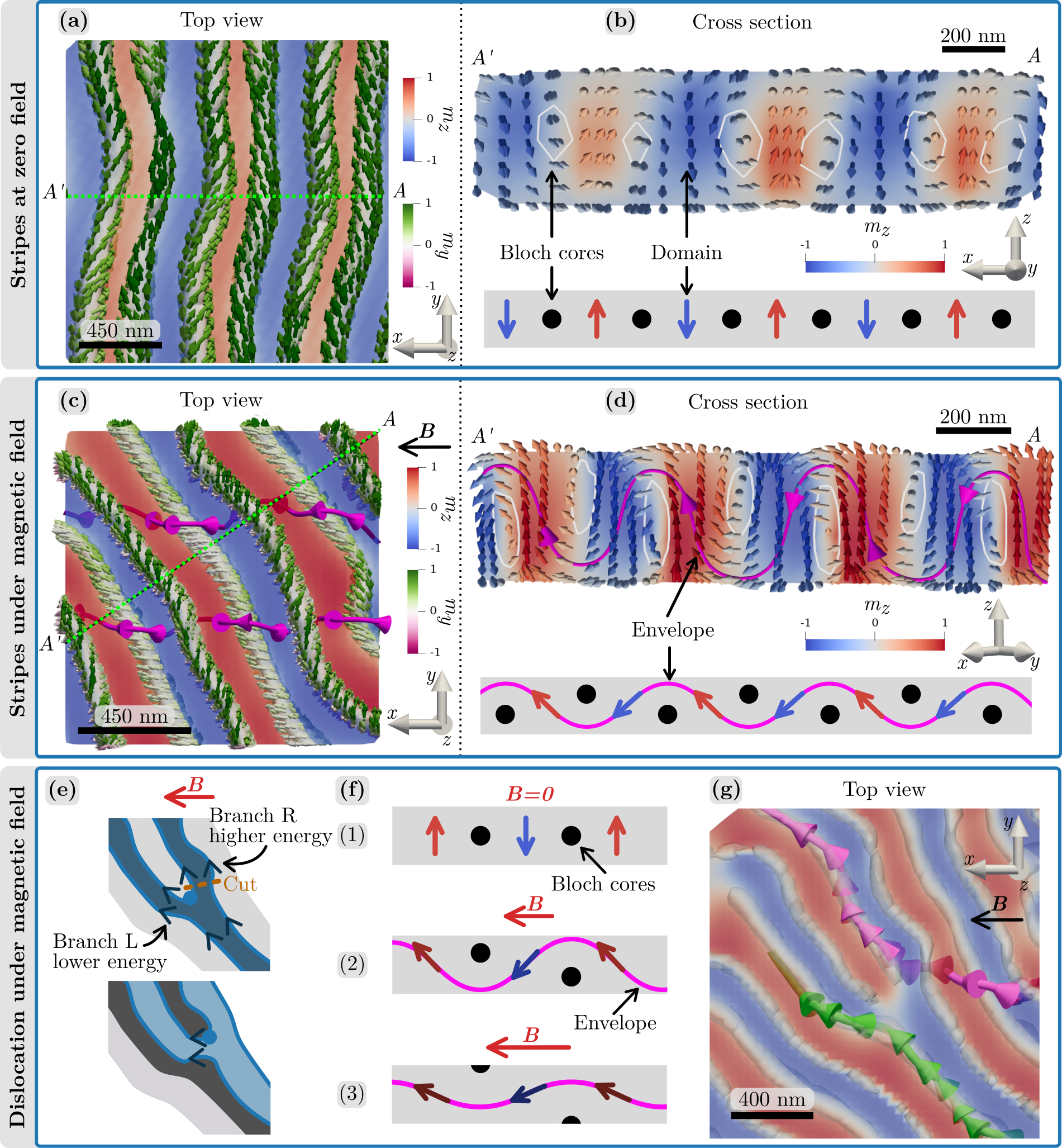}
\end{center}
\caption{Effect of the magnetic field on the 3D magnetic configuration of the stripe domains measured with X-ray laminography: Top views (a,c,g) show in red-blue contrast the out-of-plane magnetization $m_z$ in a section through the middle of the sample, and the full vector field in the Bloch cores. The Bloch cores are estimated as isosurfaces of higher vorticity with respect to the rest of the magnetic structure (grey tubes). These tubes appear as white circles in the cross sections (b,d). When a magnetic field $B=16$\,mT is applied, the stripes rotate. The in-plane magnetization also rotates and, as a result, an envelope of magnetization (pink streamlines) passing over and under the Bloch cores is formed, expelling the Bloch cores towards the surface of the sample. 
% This envelope appears when considering streamlines that represent the flow of the magnetization. 
Near the dislocations (e), the upper illustration shows the in-plane magnetization in the domain walls of a dislocation. The cut on the right (R) is facilitated by the envelope as illustrated in (f). The lower illustration in (e) represents the final state after the branch R is cut, resulting in the dislocation (shaded in blue) moving upwards. (g) The experimental configuration confirms the presence of the envelope in the right branch.}
\label{fig:fig4}
\end{figure*}

% zero field - > magnetic configuration described
The 3D magnetic configuration of the stripes at zero field, shown in Figs.~\ref{fig:fig4} (a) and (b), reveals out-of-plane domains (red and blue) and vortex-like domain walls with a core that points along the direction of the domain wall, known as Bloch cores. These Bloch cores are all oriented in the same direction (as seen by the $m_y$ component in Fig.~\ref{fig:fig4} (a)), resulting in a net in-plane magnetization. 
%typical of films with rotatable anisotropy.
%As a magnetic field perpendicular to the initial stripe orientation is applied, the in-plane magnetization aligns more strongly with the field, reducing the Zeeman energy, and resulting in the stripes rotating counterclockwise, as shown in Fig.~\ref{fig:fig4} (c).
As a magnetic field perpendicular to the initial stripe orientation is applied,
this in-plane magnetization prefers to align with the field to reduce the Zeeman energy. This breaks the symmetry between the rotation of the stripes clockwise and counterclockwise, promoting a counter clockwise rotation as seen in our experimental 2D images.%  Fig.~\ref{fig:fig4} (c), and in the 2D experiments.

% field - > change in the 3D magnetic config
When the 3D configuration under the application of field is mapped in Fig.~\ref{fig:fig4} (c) and (d), we observe that the field shifts the cores of the domain walls towards the surface, as predicted in~\cite{tacchi2014,fin2015}, resulting in a net undulating flow of the magnetization in the direction of the field that is not present at zero field, referred to here as the ``envelope'' (dark-pink streamlines).

% field - > dislocations
To understand the effect of this change in the 3D magnetic configuration on the movement of the defects, we consider the magnetization in the domain walls in the vicinity of a negative dislocation, as shown in Fig.~\ref{fig:fig4} (e).
Here the magnetization of the domain walls in the dislocation's right branch is  less aligned to the magnetic field than that of the left branch, meaning that the right branch has higher energy than the left one.
In order for the dislocation to move, one of the branches must break which, due to its higher energy, is more likely to be the right branch. This would result in the upward-branching dislocation moving in the opposite direction of the field, akin to a negatively charged particle, as observed in Fig.~\ref{fig:fig2}(d).
Conversely, downward-branching dislocations move towards the direction of the field, resembling positively charged particles.

% field - > how the dislocations move/ break
Having determined the cause of the sense of rotation of the stripes and the sense of propagation of the dislocations, we next delve deeper into the underlying mechanism of their motion, specifically focusing on the breaking of the branch.
% When we schematically consider the shifting of the Bloch cores, the resulting envelope in the magnetisation has an effect on the dislocations. 
We consider the right branch and take a cross section near the core (orange line in Fig.~\ref{fig:fig4} (e)). In Fig.~\ref{fig:fig4} (f), we represent schematically the magnetic structure of this cross section and illustrate how the closure domains, Bloch cores, and the envelope may evolve as the field increases.
At zero field $B=0$ (1), up and down magnetic domains are separated by Bloch cores (black circles).
As $B$ increases, the global state becomes energetically unfavorable, leading the magnetization to begin to align along the field (2).
Consequently, the magnetization locally rotates into the plane forming an envelope (pink line) that pushes the Bloch cores towards the surface.
With higher fields, the magnetization aligns more strongly with the field, straightening the envelope as if it were a rope (3).
While this phenomenon occurs globally, in the vicinity of the dislocations, the in-plane tilting acts to expel two domain walls and merge the domains, as shown in Fig.~\ref{fig:fig4} (e), effectively breaking the branch and shifting the dislocation.
This behavior is confirmed by the mapping of the 3D magnetisation field in the vicinity of the dislocation, where we indeed observe an envelope forming across the right branch of a negative dislocation shown in Fig.~\ref{fig:fig4} (g), as in Fig.~\ref{fig:fig4} (f)(2). This effectively guides the dislocation motion and facilitates the rotation of the stripes.
By visualising the 3D configuration, we gain a deeper understanding of the rotation of the stripes and the motion of the defects, highlighting the multidimensional nature of the system. Specifically, the underlying 3D structure and its response to magnetic fields reveals the mechanism of motion of dislocations and their charged-particle nature.

\section{Modelling defect motion and stripe rotation}

\begin{figure*}[!t]
\begin{center}
\includegraphics[width=0.9\linewidth]{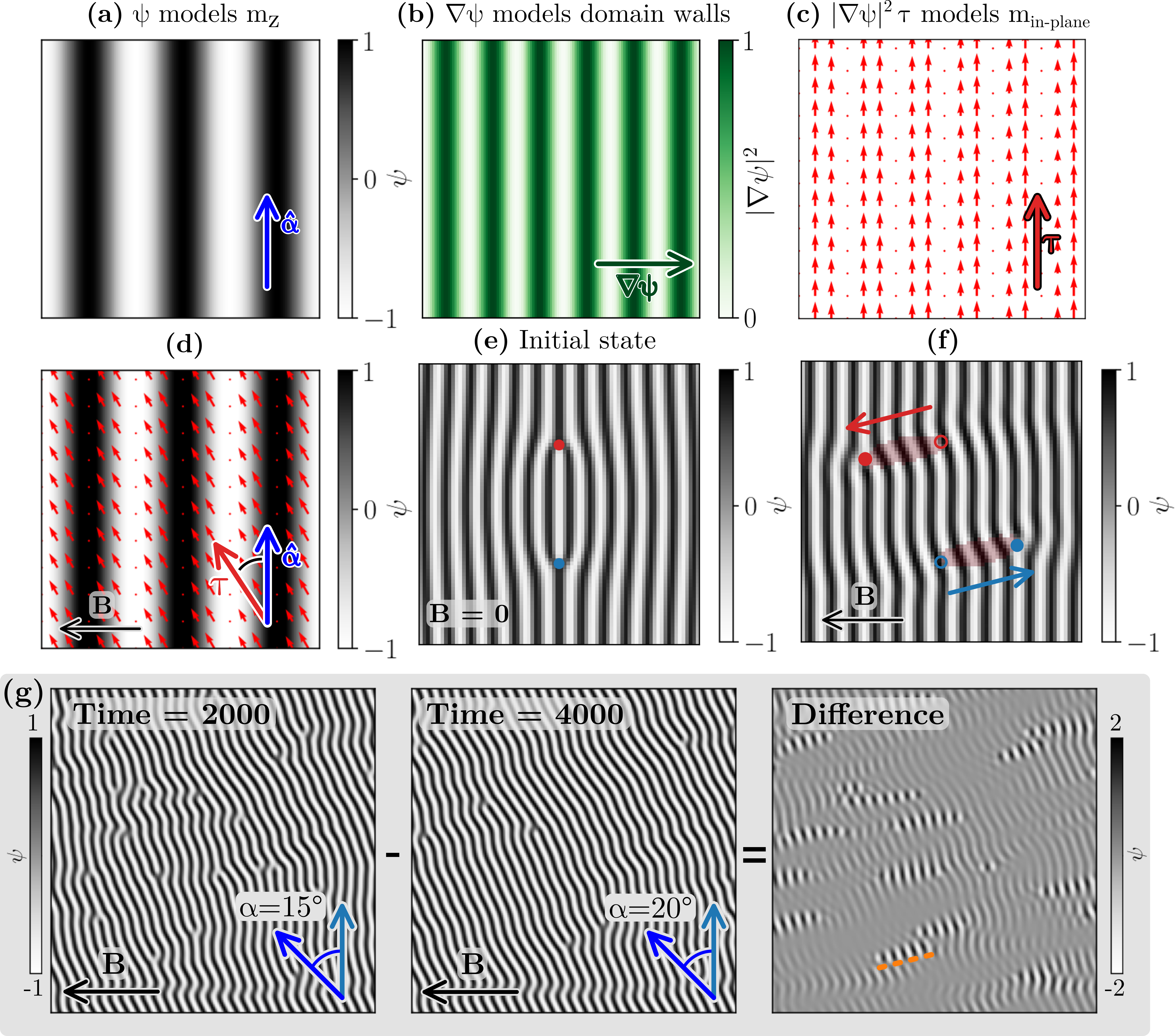}
\end{center}
\caption{Modeling unidirectional motion and continuous stripe rotation:
(a) Stripe pattern described by $\psi$ as minimizer of the free energy, representing the out-of-plane magnetization.
(b) The term $|\nabla \psi|^2$ mimics the domain walls.
(c) To account for the in-plane magnetization $m_{\mbox{in-plane}}$, we introduce the vector $\bm{\tau}$.
(d) A setting featuring the deviation of $\boldsymbol{\tau}$ from the direction of (unperturbed) stripes due to the action of the magnetic field $\mathbf{B}$.
(e) Introducing a pair of positive (red) and negative (blue) dislocations disrupts the ideal stripe ordering.
(f) The external field induces motion of the defects, which also allows for rotation of stripes (see shaded red regions).
(g) Adding more dislocations reveals a difference image similar to the experiments where the lines correspond to dislocation paths.
}
\label{fig:fig5}
\end{figure*}

Having understood the 3D mechanism by which dislocations move, we now focus on the universality of this phenomenon. Specifically, we consider a minimal approach
based on the Swift-Hohenberg (SH) model~\cite{swift1977hydrodynamic,hohenberg1977theory} - or non-conservative phase field crystal (PFC) \cite{Elder2002,Emmerich2012} - supporting the modeling of stripe phases and dislocations. This approach builds on a free energy density
\begin{equation}
f_{\mbox{SH}} = \frac{1}{4}  \psi^4 - \frac{1}{2} \epsilon \psi^2 + \frac{1}{2} \kappa \psi (q_0^2 + \nabla^2)^2 \psi ,
\label{eq:model}
\end{equation}
where the smooth order parameter $\psi\equiv \psi(\mathbf{r})$ mimics the out-of-plane magnetization component $m_z$. The parameter $\epsilon$, which controls the maximum value of $\psi$, is set to $0.5$ to represent fully out-of-plane magnetization, typical of magnetic systems with perpendicular anisotropy.
The last term promotes the formation of stripes with a periodicity of $2\pi/q_0$~\cite{Emmerich2012}, as shown in Fig.~\ref{fig:fig5}~(a) (see Methods for more details).

Although the SH model successfully leads to the presence of stripe phases reminiscent of our weak magnetic stripes, to model the magnetic field induced motion of the stripes, we introduce further terms into the free energy that consider the net in-plane direction of the domain walls. 
Specifically, we introduce the unit vector
$\boldsymbol \tau\equiv \boldsymbol \tau(\theta)=[-\sin(\theta),\cos(\theta)]$ with $\theta$ an additional model variable that represents the direction of the in-plane magnetization. This direction is modulated by $|\nabla \psi|^2$ to reflect the enhanced in-plane magnetization in the domain walls (see Fig.~\ref{fig:fig5} (b) and (c)). 
Our model is then described by the free energy $F[\psi,\theta]=\int_\Omega f {\rm d}\mathbf{r}$ with
\begin{equation}
    f = \underbrace{f_{\mbox{SH}}}_{\text{Stripes}}
    - \underbrace{\frac{1}{2} |\nabla \psi|^2 \bm{\tau} \cdot \mathbf{B}}_{\text{Zeeman}} + \underbrace{\frac{\gamma}{2} \left( \boldsymbol{\tau} \cdot \nabla \psi \right)^2}_{\text{Bloch-type walls}} + \underbrace{\sigma |\nabla \theta|^2}_{\text{Exchange-like}}.
\label{eq:model2}
\end{equation}
The second term mimics the Zeeman energy and is minimized when
$|\nabla \psi|^2\boldsymbol{\tau}$ is parallel to the magnetic field $\mathbf{B}$. The third term promotes the Bloch-type nature of the domain walls, coupling $\boldsymbol{\tau}$ to the direction parallel to the stripes (perpendicular to $\nabla \psi$). 
% We note that these terms resemble lowest order contributions in SH/PFC formulations encoding magnetization and the effect of external magnetic fields in crystalline materials \cite{Faghihi2013,Seymour15,Backofen2019}. 
The last term avoids sharp gradients in the orientation of $\boldsymbol{\tau}$, similar to an exchange energy within the domain. 
To model the rotation, the state is allowed to dynamically relax by evaluating non-conservative coupled gradient flow equations as described in the Methods.
%To model the rotation, the dynamics of out-of-equilibrium settings is evaluated via a non-conservative, coupled gradient flow equations for $\psi$ and $\theta$, namely $\partial_t \psi=-\delta F[\psi,\theta]/\delta \psi$ and $\partial_t \theta=-\delta F[\psi,\theta]/\delta \theta$ (for further details, see the section Methods).
% Further details concerning the full expression of these equations, parameters, the adopted numerical scheme, and the initialization of patterns with dislocations are reported in the Method Section.

In the absence of an external magnetic field ($\mathbf{B}=\mathbf{0}$), the Bloch-type walls energy terms enforce the in-plane magnetization modeled by $\tau$ to be aligned with the stripes, which are perpendicular to $\nabla \psi$ (see Fig.~\ref{fig:fig5} (a,b,c)). When increasing $\mathbf{B}$ the Zeeman energy would favor the alignment of the in-plane magnetization to $\mathbf{B}$, consequently favoring the alignment of the stripes themselves to $\mathbf{B}$.
We find, however, that the lack perturbations in the system hinders the rotation of stripes, as shown in Fig.~\ref{fig:fig5} (d).

The importance of defects in mediating the rotation of the stripes becomes clear when they are introduced into our minimal model.
The pair of positive and negative dislocations shown in Fig.~\ref{fig:fig5} (e) create perturbations in the orientation of the stripes. 
These perturbations result in a symmetry breaking when an external field $\mathbf{B}$ is applied, which leads to the motion of the defects, as shown in Fig.~\ref{fig:fig5} (f), and causes the stripes to rotate as the dislocations move.
The positive (negative) dislocation moves to the left (right), in the (opposite) direction of $\mathbf{B}$, resulting in the counterclockwise rotation of the stripes, as observed in our 2D experiments (Fig.~\ref{fig:fig2} (c,d)). This movement of the dislocations enables the alignment of the in-plane component with $\mathbf{B}$, as observed in the 3D experiments (Fig.~\ref{fig:fig4} (c,d)).

Having determined that the defects are responsible for the rotation of the stripes, we next consider a system with a higher number of defects, as in the experiments, and we calculate the difference image between consecutive configurations, as shown in Fig.~\ref{fig:fig5} (g).
This image reveals clear paths of unidirectional motion for the dislocations as observed experimentally.
% We note that
% , although the symmetry breaking and the overall behavior of dislocations with opposite topological charges match qualitatively the experimental observation, 
% the motion occurs along a trajectory that is at a lower angle with respect to the one observed experimentally.
When comparing Fig.~\ref{fig:fig5} (g) to Fig.~\ref{fig:fig2} (a), we observe a slight difference in the angle of the 1D motion with respect to the experiments which may be due to the more complex 3D magnetic configuration that is not considered in the model. 
Indeed, the model considers only the simple 2D configuration, meaning that the observed results - the 1D motion, dislocation-mediated stripe rotation and the charged particle behaviour - are not limited to our magnetic stripe system, but can be considered in the context of a much wider variety of stripe hosting systems.

With analogous phenomena in surface wrinkles~\cite{ohzono2006defect}, polymer assembly~\cite{hur2018defect,kim2019shear}, and grain rotation in polycrystalline materials~\cite{wang2014grain,tian2024grain},
the motion of defects mediating continuous transitions plays an important role in the physics of a number of systems. 
By being able to reproduce the 1D motion of charged particles within a magnetic system with a minimal model, this work paves the way to a universal understanding of dislocation motion in striped systems influenced by in-plane fields. This understanding will inform a number of phenomena, from mechanical properties of materials, to the self assembly of polymers, as well as information carrying defects in magnetic systems.

% Overall, our minimal model retaining stripe ordering, defects, and coupling with an externally induced preferential orientation, successfully reproduces the key experimental features: 
% (1) dislocations mediate stripe rotation,\
% (2) they act like charged particles moving in opposite directions, and 
% (3) their paths exhibit a well-defined 1D direction.
% Going beyond magnetism, analogous phenomena of defect-driven rotations have been observed across a wide variety of systems, from surface wrinkles~\cite{ohzono2006defect} and polymer assembly directed by substrates~\cite{hur2018defect,kim2019shear}, to grain rotation in polycrystalline materials~\cite{wang2014grain,tian2024grain}.
% We thus obtain analogous evidence for systems featuring magnetic stripes. At the same time, the proposed formulation paves the way to a universal understanding of dislocation motion in striped systems influenced by in-plane fields.

%%%%%%%%%%%%%%%%%%%%%%%%
\section{Discussion and outlook}
\label{sec:conclusions}

In conclusion, we demonstrate the unidirectional propagation of topological defects in a laterally unconfined system that mediates a continuous transition of the order parameter, where the defects can be propagated in different directions in a 2D plane. 
Using a comprehensive approach that combines 2D and 3D X-ray magnetic imaging with phenomenological modeling, we reveal that the motion of magnetic dislocations is multidimensional: their one-dimensional movement facilitates a two-dimensional rotation of magnetic stripes, driven by the three-dimensional magnetic configuration.

Our results show that magnetic dislocations rotate the stripes locally as they move, enabling the system to gradually adjust its average orientation in response to an external magnetic field. We find that in-plane magnetization, characteristic of weak magnetic stripes, dictates the positive or negative charge behavior of the dislocations. This robust motion, reproducible in a simple phenomenological model, enhances our understanding of topological defects in unconfined systems and opens avenues for controlled defect motion beyond nanostructured materials.

From a technical standpoint, we establish field-dependent X-ray magnetic laminography by developing a dedicated sample holder with stackable permanent magnets, which allows for a tunable magnetic field, up to $100$\,mT in magnitude, and full vectorial reconstruction without disrupting the magnetic configuration.
This capability opens new possibilities to study the field-driven evolution of magnetic configurations, to explore a broader range of materials and phenomena.

Regarding the behavior of our model system, here we highlight the ability to stabilize and manipulate the orientation of magnetic stripes.
The dense presence of magnetic defects allows for a smooth transition in stripe orientation from 0 to $90$\textdegree{}.
Previous studies have demonstrated such analogue behavior in helimagnets - also known as helitronics~\cite{bechler2023helitronics}. However, in that case, the helical stripes are limited to vertical or horizontal orientations, meaning that analogue values between 0 and 1 are obtained from the percentage of vertical stripes compared to horizontal stripes. In contrast, in our system the angle of the stripe orientation can be treated as a continuous, analogue quantity.
Notably, upon removal of the external magnetic field, the orientation of the stripes remains stable, enabling the encoding of non-volatile analogue information - potentially relevant for neuromorphic computing~\cite{sander20172017,olejnik2017antiferromagnetic}.

Lastly, we note that the behavior of the magnetic dislocations in this system resembles that of fractons, pseudo-particles that move in 1D within a 2D system.
This controlled 1D motion in Py thin films offers a platform to explore fractionalized excitations and their restricted mobility in a well established magnetic system.
Key to exploiting this behavior will be the manipulation of dislocation propagation angles, which may be achieved by inducing specific perpendicular magnetic anisotropy through tuning the film's growth parameters.
The realization of such unidirectional behavior of topological defects paves the way for further exploration into the intriguing dynamics and applications of fractons within quantum condensed matter systems.

\section{Methods}
\label{sec:methods}
\subsection{Sample preparation}
Permalloy (Py) films were grown at room temperature in an ultra-high vacuum magnetron sputtering deposition system, which was evacuated to a base pressure lower than $5\times10^{-9}$\,mbar prior to the growth. For the growth of Py, the partial pressure of Argon gas was set specifically to $9\times10^{-3}$\,mbar, to promote the magnetic stripe order of Py not achievable at lower Ar working pressure. The target-to-substrate distance was fixed to $20$\,cm, and the substrate rotated at $24$\,rpm to ensure a spatially homogeneous sample thickness and composition. We used a $7.62$\,cm in diameter Ni$_{81}$Fe$_{19}$ target, in a face-to-face geometry, and applied a continuous (DC) source power of $40$\,W, corresponding to a deposition rate of $0.47$\,\AA s$^{-1}$. The films were capped \textit{in situ} by $4$\,nm of Au, sourcing $50$\,W DC power to a $7.62$\,cm in diameter target, at a working pressure of $3\times10^{-3}$ mbar of Ar, corresponding to a deposition rate of $1.33$\,\AA s$^{-1}$.

\subsection{STXM and Laminography}
The scanning transmission X-ray microscopy and laminography experiments were performed at the PolLux (X07DA) endstation of the Swiss Light Source.
The photon energy was tuned to $722.4$\,eV, which corresponded to the maximum X-ray magnetic circular dichroism signal at the Fe $L_2$ edge.

We acquired $30$ projections around the laminography axis, which in turn was tilted $45$\textdegree{} with respect to the X-ray beam direction.
The sample was rotated by a regular angular step of $12$\textdegree{} around the $360$\textdegree{}.
Each projection image was measured $3$\,ms of exposure time in a field of view of $10\times10$\,$\upmu$m$^2$ with $250\times250$ points, giving a pixel size of $40$\,nm.

A special sample holder was designed for the \textit{in situ} application of the external magnetic field during the laminography measurement. This sample holder has space at both sides of the sample for stacking a number of small permanent magnets. The size of each magnet is $1\times1\times5$\,mm$^3$ and they create an in-plane magnetic field in the center of the sample holder of approximately $5$\,mT each. We calibrated the intensity of the field without the sample with a Hall probe before starting the experiment. Since they are embedded in the sample holder and they do not protrude, they do not interfere with the beam nor any other part of the experimental setup.

\subsection{Minimal model}
%To model the rotation, the dynamics of out-of-equilibrium settings is evaluated via a non-conservative, coupled gradient flow equations for $\psi$ and $\theta$, namely $\partial_t \psi=-\delta F[\psi,\theta]/\delta \psi$ and $\partial_t \theta=-\delta F[\psi,\theta]/\delta \theta$ (for further details, see the section Methods).
The model presented in the main text results in two dynamical equations, corresponding to non-conservative gradient flows for $\psi$ and $\theta$ 
\begin{equation}
\begin{split}
\partial_t \psi=-\frac{\delta \mathcal{F}}{\delta \psi} =&\epsilon \psi - \psi^3 - \kappa (1+\nabla^2)^2 \psi \\ & + \gamma \nabla \cdot [(\boldsymbol{\tau} \cdot \nabla \psi)\boldsymbol{\tau} ] - \nabla \cdot [ (\boldsymbol{\tau} \cdot \mathbf{B}) \nabla \psi], \\
\partial_t \theta = -\frac{\delta \mathcal{F}}{\delta \theta} 
=& \sigma \nabla^2 \theta - \frac{1}{2}\mathbf{n} \cdot \mathbf{B} |\nabla \psi|^2
\\ &+ \gamma (\mathbf{n} \cdot \nabla \psi)(\boldsymbol{\tau} \cdot \nabla \psi),
\label{eq:dndt}
\end{split}
\end{equation}
with $\mathbf{n}\equiv{\mathbf{n}}(\theta)={\boldsymbol{\tau}}(\theta)^\prime$.
These equations can be numerically solved by a standard Fourier pseudo-spectral method. By rewriting the equation above as $\partial_t{\psi} = \mathcal L \psi + \mathcal N (\psi)$ with $\mathcal{L}$ and $\mathcal{N}$ the linear and nonlinear operators, we solve for $
\partial_t\widehat{\psi} = \widehat{\mathcal L}_k [\widehat{\psi}]_k + \widehat{\mathcal N}_k$ with $[\widehat{\psi}]_k$ the coefficient of the Fourier transform of $\psi$, $\widehat{N}_k$ the Fourier transform of $N(\psi)$ and $\widehat{L}_k$ the Fourier transform of $\mathcal{L}$ resulting in an algebraic expression of the wave vector. The solution at $t+\Delta t$, with $\Delta t$ the time step, is then obtained via an inverse Fourier transform of $[\widehat{\psi}_n]_k(t+\Delta t)$ computed by the following approximation
\begin{equation}
[\widehat{\psi}]_k(t+\Delta t) = \frac{[\widehat{\psi}]_k(t) + \Delta t \widehat{\mathcal N}_k(t)}{1-\Delta t \widehat{\mathcal L}_k}.
\label{eq:spec}
\end{equation}
Solutions $\psi(t)$ are obtained by an inverse Fourier transform of $[\widehat{\psi}_n]_k(t)$. We note that this method naturally enforces periodic boundary conditions.

The parameters selected for the simulations reported in Fig.~\ref{fig:fig5} are the following: $q_0=1$, $\epsilon = 0.5$, $\kappa = 1$, $\gamma=2$, $\sigma=0.1$ and $B=0.5$, with the time and spacial discretizations $\Delta t=0.1$ and $\Delta x=1.25$, respectively.

An unperturbed stripe phase can be initialized by $\psi_0=\eta e^{i \mathbf{k}\cdot \mathbf{r}}+ \bar \psi$ with $\mathbf{k}$ the principal wave vector and $\bar \psi$ the mean of $\psi_0$. To consider vertical stripes, as observed in the experiments, we initially set $\mathbf{k}=[q_0,0]$ and $\eta= (2/3) \sqrt{3\epsilon-9\bar \psi^2}$ minimizing the energy for $\mathbf{B}=\mathbf{0}$ (we set here $\bar \psi=0$). We note that non-zero external fields slightly vary the equilibrium wave number; however, this effect does not affect the evidence discussed above.

A smooth field $\bar \psi$ hosting dislocations can then be initialized by imposing $\psi=\psi_0 e^{-i \mathbf{k}\cdot \mathbf{u}}$ with $\mathbf{u}^{\rm dislo}$ the displacement field induced by a dislocation~\cite{salvalaglio2022coarse}. For a dislocation in 2D with Burgers vector aligned along the $x$-direction with length $b$ it reads~\cite{anderson2017}
\begin{equation}
    \begin{split}
        u_x^{\rm dislo}&= \frac{b}{2\pi} \bigg[ \arctan{\left(\frac{y}{x}\right)} +\frac{xy}{2(1-\nu)(x^2+y^2)} \bigg],\\
        u_y^{\rm dislo}&= -\frac{b}{2\pi} \bigg[ \frac{(1-2\nu)}{4(1-\nu)}\log{\left( x^2+y^2 \right)}+\frac{x^2-y^2}{4 (1-\nu) (x^2+y^2)} \bigg],\\
    \end{split}
    \label{eq:udislo}
\end{equation}
Dislocations are initialized with $b=\pm 2\pi$. $\nu$, namely the Poisson ratio, is set to $1/3$ although the elastic constant encoded in the energy functional is then obtained when relaxing the initial condition via the gradient flow Eq.~\eqref{eq:dndt}~\cite{BenoitMarechal_2024}.

\section*{Acknowledgments}
The 3D magnetic laminography was performed at the PolLux (X07DA) endstation of the Swiss Light Source, Paul Scherrer Institut, Villigen PSI, Switzerland. The PolLux end station was financed by the German Ministerium für Bildung und Forschung (BMBF) through contracts 05K16WED and 05K19WE2.
M. Di Pietro Martínez, L. A. Turnbull, J. Neethirajan and C. Donnelly acknowledge funding from the Max Planck Society Lise Meitner Excellence Program and funding from the European Research Council (ERC) under the ERC Starting Grant No. 3DNANOQUANT 101116043.
M. Di Pietro Martínez and L. A. Turnbull acknowledge the support of the Alexander von Humboldt Foundation.
M. Salvalaglio acknowledges the support of the Emmy Noether Programme of the Deutsche Forschungsgemeinschaft (DFG, German Research Foundation) project No. 447241406.
A. Hierro-Rodríguez and M. Vélez acknowledge support from Spanish MCIN/AEI/10.13039/501100011033/FEDER,UE under grant PID2022-136784NB.
The authors acknowledge Markus König and Renate Hempel-Weber for their assistance during FIB and VSM-SQUID experiments, respectively. 
The authors thank Roderich Moessner and Frank Pollmann for fruitful discussions, their careful reading and insightful comments on this manuscript.

\bibliography{rotdis.bib}

\end{document}